\documentclass[aip,reprint]{revtex4-2}

\usepackage{algpseudocode}
\usepackage{chemformula} 
\usepackage[T1]{fontenc} 
\usepackage{graphicx}
\usepackage{bm}
\usepackage{hyperref}
\hypersetup{colorlinks = true, urlcolor = blue, linkcolor = blue, citecolor = blue}
\usepackage{color}
\usepackage[normalem]{ulem}
\usepackage{float}
\usepackage{subfigure}
\usepackage{amsmath}
\usepackage{graphicx}
\usepackage{amsfonts}
\usepackage{mathrsfs,relsize,array}
\usepackage{appendix}
\usepackage{romannum}
\usepackage{makecell}
\usepackage{siunitx}

\let\vec\mathbf

\begin{document}


\title{Role of a fractal shape of the inclusions on acoustic attenuation in a nanocomposite} 



\author{H. Luo}
\affiliation{ 
Univ Lyon, INSA-Lyon, CNRS UMR5259, LaMCoS, F-69621, France
}%

\author{Y. Ren}
\affiliation{ 
State Key Laboratory of Solidification Processing, Center of Advanced Lubrication and Seal Materials, Northwestern Polytechnical University, Xi'an, Shaanxi710072, PR China
}%

\author{A. Gravouil}
\affiliation{ 
Univ Lyon, INSA-Lyon, CNRS UMR5259, LaMCoS, F-69621, France
}%

\author{V. M. Giordano}
\affiliation{ 
Institut Lumi\`ere  Mati\`ere, UMR 5306 Universit\'e Lyon 1-CNRS,  F-69622 Villeurbanne Cedex, France
}%

\author{Q. Zhou}
\affiliation{ 
State Key Laboratory of Solidification Processing, Center of Advanced Lubrication and Seal Materials, Northwestern Polytechnical University, Xi'an, Shaanxi710072, PR China
}%

\author{H. Wang}
\affiliation{ 
State Key Laboratory of Solidification Processing, Center of Advanced Lubrication and Seal Materials, Northwestern Polytechnical University, Xi'an, Shaanxi710072, PR China
}%

\author{A. Tanguy}
\email{anne.tanguy@insa-lyon.fr.}
\affiliation{ 
Univ Lyon, INSA-Lyon, CNRS UMR5259, LaMCoS, F-69621, France
}%
 \affiliation{ONERA, University Paris-Saclay, Chemin de la Hunière, BP 80100, 92123, Palaiseau, France.}

\date{\today}

\begin{abstract}
Nanophononic materials are promising to control the transport of sound  in the GHz range and  heat in the THz range. 
Here we are interested in the influence of a dendritic shape of inclusion on acoustic attenuation. 
We investigate a Finite Element numerical simulation of the transient propagation of an acoustic wave-packet 
in 2D nanophononic materials with circular or dendritic inclusions periodically distributed in matrix.
By measuring the penetration length, diffusivity, and instantaneous wave velocity, we find that the multi-branching tree-like form of dendrites provides a continuous source of phonon-interface scattering leading to
an increasing acoustic attenuation. When the wavelength is far less than the inter-inclusion distance, we report a strong attenuation 
process in the dendritic case which can be fitted by a compressed exponential function with $\beta>1$.

\end{abstract}


\maketitle 


\section{Introduction}



Heterogeneous architectured materials are of particular interest in engineering applications \cite{Bouaziz2008}.
 These are manmade structural
materials which have been developed for obtaining ad hoc
properties, that cannot be generally found in nature. They
are  obtained by engineering at different scales the
mixing of different materials, either in a random spacial
distribution (composite materials) either with the artificial repetition of regular patterns (metamaterials). Depending on their application, the lengthscale of such patterns can span from the nanometer to the macroscopic range, being smaller than, or comparable to, the wavelength of the phenomena that the material is meant to affect.

This concept has indeed been largely exploited for efficiently manipulating long-wavelength acoustic phonons, which assure sound propagation at low frequency, through the introduction in materials of periodic interfaces on a macroscopic scale (phononic crystal, PC) \cite{Economou1989,Maslov1999,Elford2010,Verdier2018}. In recent years, they have
attracted increasing attention among the scientific community
due to their extraordinary acoustic and elastic wave propagation
performances obtained by designing the phase gradient at the sub-wavelength scale,  such as negative refraction, waveguiding, cloaking, and band gaps ~\cite{Liu2000,Khelif2006,Zhang2009,Cummer2016}. 

Thanks to both engineering
and theoretical progresses, they have been introduced also
in thermal science, with a microstructure in the nanometer
scale.
Thermal transport  is  intimately related to the sound propagation (acoustic transfer) in materials because in insulators and semi-conductors the main heat carriers are acoustic phonons~\cite{Vandersande1986}.
Specifically, at room temperature heat is mainly carried by phonons with THz frequencies and nanometric wavelengths. As such, a periodic nanostructuration has proved to be promising for affecting phonon dispersions and their ability in transporting heat \cite{Hussein2019}. 
It is important to remind here that phonons participate to thermal transport through  two different contributions: a propagative one, which depends on phonon mean-free path, heat capacity, velocity and vibrational density of states~\cite{Kittel2004}, and~a diffusive one, involving the phonon diffusivity rather than the mean-free path and velocity~\cite{Larkin2014,Allen1990,Allen1999}, this latter dominating at higher energies and smaller wavelengths. 
The propagative contribution can be reduced through the presence of interfaces, which scatter or eventually trap phonons. Depending on the relative importance of the diffusive contribution, in fact, the reduction but also the improvement of thermal conductivity have been reported in different nanostructured  systems~\cite{Choi1995,Wang2013,Schlichting2001}.
Numerous atomistic simulations of out-of-equilibrium phonon transport  have also been reported in different systems ~\cite{Prasher2009,Termentzidis2009,Merabia2012,Zen2014,France-Lanord2014,Moon2016,Anufriev2016,Tlili2017,Tlili2019}. 
Recent works have looked at the phonon dynamics to get a better insight on transport properties and found exotic behaviors such as an energy localization between pores~\cite{Morthomas2019}, asymmetric transport (rectification) \cite{Desmarchelier2021} or the filtering of high frequency phonons~\cite{Damart2015}. 

It is clear thus, that, depending on the lengthscale at play, the design of the phononic crystals structure contributes directly to the performance of filtering, hindering, and guiding the propagation of acoustic waves (phonons), responsible for the sound propagation when their wavelength is macroscopic, and for thermal transport at room temperature when is nanometric \cite{Tanaka2000,France-Lanord2014,Damart2015,Anufriev2017,Meyer2016}. 

Many theoretical studies have tried to shed light onto  the key parameters of the phononic crystals that determine the  effect of the interfaces on acoustic and thermal transport. It is worth mentioning that researches on mechanical  and geometrical properties of the nano-interfaces are at the core of the design of the structured materials, involving, for example,  the shape and dimensions of the inclusion, the contrast of properties between the materials on the two sides of the interfaces. Recently, we have shown that circular interfaces in a 2D nanophononic material affect differently phonons of different wavelengths  and it is necessary to look at all phonons relevant for heat transport at a certain temperature, and their perturbed dynamics, for being able to understand thermal conductivity in such nanocomposites~\cite{Tlili2017,Tlili2019}.
In that work we could highlight that the rigidity contrast between the two phases is a determinant parameter controlling the strength of the scattering which affects the phonons with a wavelength comparable with the nanostructuration lengthscale, significantly anticipating the propagative-to-diffusive crossover in an amorphous matrix. A better understanding of the mechanisms at play and the role of the different parameters could be acquired with a more
complete and systematic parametric study on the combined effect of rigidity contrast, interface density, nanostructure lengthscale and phonon wavelength. Our results allowed us to identify  different transfer regimes (propagative, diffusive, localized or mixed regimes) in an elastic nanophononic 2D crystal  and to show that softer inclusions are more efficient for energy attenuation, but rigid inclusions are also able to pin the vibrational energy at specific frequencies\cite{Luo2019}.  In addition, the existence of an optimal radius of circular inclusion clearly shows that, instead of monotonously increasing  the volume fraction until the percolation effect is dominant,  other geometrical parameters of the interface are relevant, such as shape, asymmetry, inclusions inter-distance and interface-to-volume ratio. Finally, when the size of the nanostructure becomes comparable to the wavelength of the excitation, various complex behaviours may occur, that affect the acoustic attenuation with a non monotonous frequency dependence. Among such phenomena are the acoustic resonances of the inclusions, multiple reflections between the inclusions, or the interfacial modes.



In contrast to most work focusing  on the periodic circular in 2D (spherical in 3D) interfaces, there are only few investigations conducted with geometrical variations.
Firstly, the specific non-aligned arrangement of circular holes (where holes can be considered as the limit of soft inclusion)
 is reported to have possibly stronger phonon attenuation than the perfectly periodic arrangement. Similarly, the emerging  of the gradient-index PCs is meant to control  independently the size of each unit hole \cite{Lin2009}. Then, recent experimental studies suggested that the non-circular holes might be more efficient in the thermal conductivity reduction and acoustic attenuation compared to conventional circular ones \cite{Schmotz2011,Gluchko2019}. 
For example, pacman-shaped holes showed a 40\% reduction of thermal conductivity compared to circular ones at room temperature, due to the high surface-to-volume ratio and possible additional resonances \cite{Gluchko2019}. 
It has been also proposed to optimize the inclusion shape in phononic structures using the homogenized model of strongly heterogeneous elastic composites\cite{Vondrejc2017}, however, this model is limited to simple shapes and topological isomorphism. 
Topology optimization of metamaterials seems to be a promising alternative\cite{Lu2013, Lee2009} but is still limited to the current level of nanotechnology manufacturing.

\begin{figure}[htbp]
\centering
\includegraphics[width=8cm]{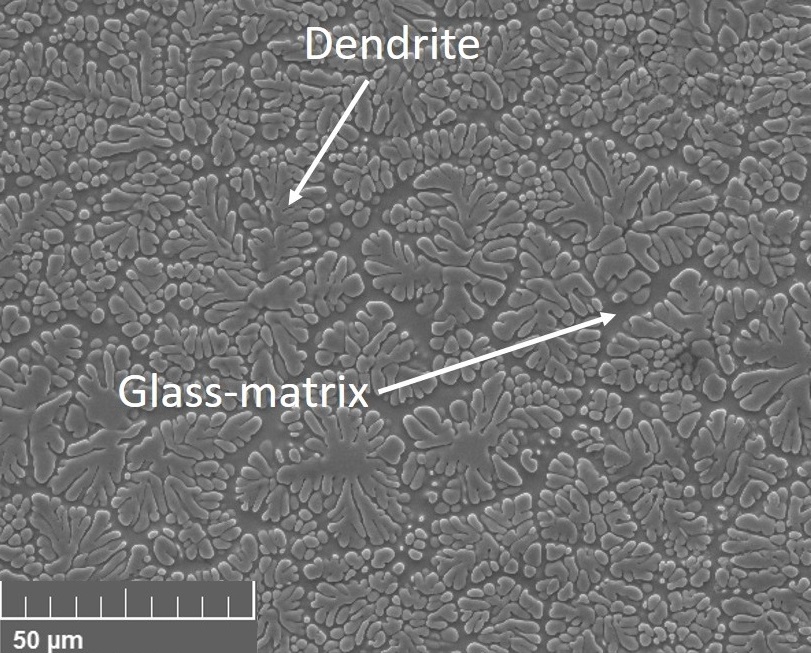}
\caption{Cross sectional SEM image of the as-cast Ti\textsubscript{45}Zr\textsubscript{25}Nb\textsubscript{6}Cu\textsubscript{5}Be\textsubscript{17}Sn\textsubscript{2}, the dendrite-phase (light gray) distributes homogeneously in the glass-matrix (dark gray). }
\label{fig:MEB_image}
\end{figure}

In this work, we propose to investigate the impact of a complex inclusion shape on acoustic propagation and attenuation at nanometric wavelengths. To this purpose, we have chosen to work on a realistic nanocomposite, as can be naturally obtained using vitrification of some metallic glasses. 
Specifically, we will work on a Ti\textsubscript{45}Zr\textsubscript{25}Nb\textsubscript{6}Cu\textsubscript{5}Be\textsubscript{17}Sn\textsubscript{2} bulk metallic glass (BMG), which, once produced by casting techniques, exhibits dendrite-phase Titanium precipitates at the micrometric scale, as shown in Fig.\ref{fig:MEB_image}.
Bulk metallic glasses  are promising structural materials  because of their excellent properties such as high yield strength, excellent corrosion resistance, and low stiffness \cite{Zhai2016,Johnson1999,Qiao2016,Inoue2011}. 
But at the same time, BMGs lack ductility and always fail in an apparently brittle manner, which seriously limits their applications \cite{Qiao2017,Schuh2007,Schroers2009,Qiao2016a}. Impressively, a larger dendrite-phase dimension offers higher ductility but decreases the yield strength of the composites. 
It has been proposed that the dendrite inclusions actually suppress the
catastrophic failure due to the propagation of shear 
bands and thus enhance the global plasticity \cite{Hays2000,Gentile2020}.
Also, the alloy compositional design could be employed to modulate the mechanical properties of the BMG, indicating a potential of tuning, for example, the stiffness ratio between the matrix and the dendrite \cite{Xu2018}.  
The multi-branches, tree-like geometry of these inclusions is of particular interest for understanding the role of a complex geometry on acoustic attenuation. 
Compared to the simple shapes, its fractal property determines a remarkably high interface-to-volume ratio, leading to potentially good  sound attenuation performance. It is thus an ideal sample for our study.

To have  a deep understanding of the role of the dendritic shape inclusion on acoustic attenuation and dynamic properties of the phononic crystal,  we have performed finite element simulations of out-of-equilibrium acoustic wave-packet propagation in 2D nanocomposites periodically distributed circular and  dendritic inclusions. This has allowed us to establish a direct comparison between simple and complex geometry.

The paper is organized as follows: in the second part,  we give a brief introduction on the  in-situ formed dendrite phase and we describe how we extract a cluster of dendritic structure from a SEM image; in the third part, we describe how the finite element calculations are performed; in the fourth part, we compare the acoustic attenuation properties in two  media containing softer or stiffer inclusions with dendritic and circular shape;
finally, we discuss the results and we conclude in the last part.

\section{The dendritic shape of the inclusions}
 
In this section, we will  present the preparation of the Ti-based metallic glass containing  dense dendrite-phase\cite{Zhai2016,Xu2018}. Based on the   SEM images of the samples, instead of looking at the whole dendrite-phase, we use a representative cluster of dendritic structure as the elementary pattern periodically distributed in our model phononic crystal. As for the mechanical properties of the model,   we have chosen to use the same properties as in Ref.\cite{Luo2019}, referring thus to a well known system. This choice is motivated by the possibility to compare wave-packet propagation results previously obtained on circular inclusions with the ones on dendritic inclusions and thus get a direct understanding of the impact of a fractal-like interface shape.

\subsection{Material composition and preparing process}

Ingots with nominal composition  Ti\textsubscript{45}Zr\textsubscript{25}Nb\textsubscript{6}Cu\textsubscript{5}Be\textsubscript{17}Sn\textsubscript{2} are prepared by arc-melting the mixture of high purity elements (>99.9wt\%) under a Ti-gettered argon atmosphere. The ingots were re-melted at least five times to ensure the homogeneity. Plate samples ($5\times 20 \times 60$, mm) are prepared by casting into a water-cooled copper mold. The dendrite phase (light gray regions) was found to distribute uniformly within the featureless glass matrix (dark gray regions). Volume fractions  of the dendrite phase are analyzed by the Image-Pro Plus software and resulting in a volume fraction $66\pm2 \%$. The dendrites have an average size (measured as the diameter of the circular approximation to its shape) of about 30 microns, with a typical single-branch diameter of 3 microns. The size and volume fraction are sensitive to the cooling rate and the alloy composition \cite{Zhang2019}. Some mechanical properties are given in the Appendix and more details on  sample preparation and materials properties  can be found in Ref.\cite{Zhai2016,Xu2018}.
 
In the following, the exact size of the inclusions will not matter much. We will keep only their shape, and their volume fraction. Thanks to the scalability of our numerical calculations, lengths will be expressed in units of $L$ (the average distance between the inclusions) and frequency in units of $\omega_0=2\pi c_L/L$ with $c_L$ the longitudinal wave velocity. The inclusions with a reported diameter of $30 ~\mu m$ are not supposed to affect the thermal transport at room temperature, but only acoustic wave propagation at GHz frequencies. But, playing with the exact value of $L$, it will be possible to consider also the effect of the inclusion shape on  thermal transport: for example by choosing  $L\approx nm$, effect on thermal transport at room temperature can be investigated.

\subsection{Reconstruction of the dendritic shape inclusion}
\begin{figure*}[htbp]
\centering
\includegraphics[width=14cm]{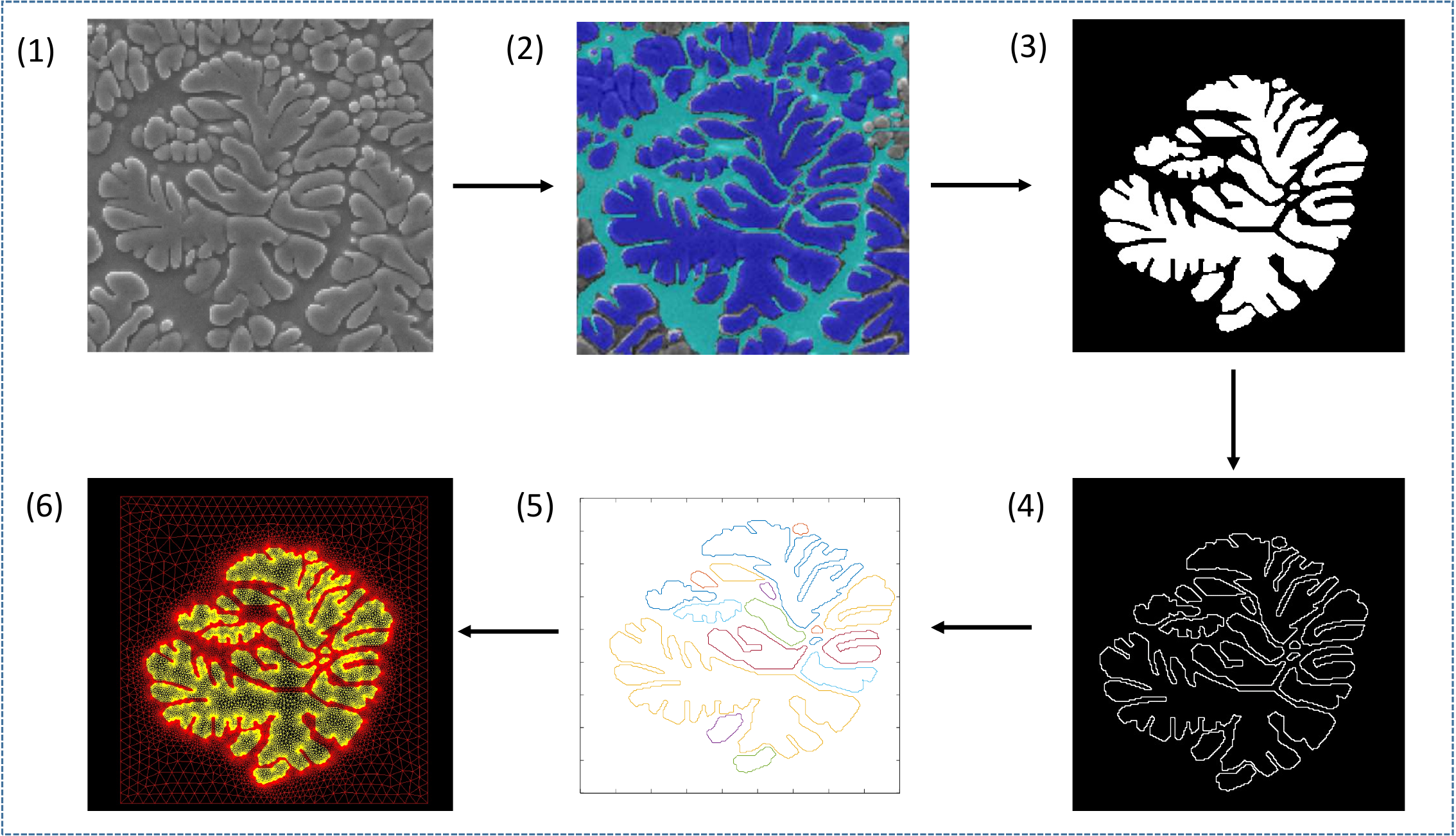}
  \caption{Flowchart from SEM to the finite element model: (1) Region of interest extracted from SEM image; (2) Pixel-level labeling using Matlab toolbox Image Labeler; (3) Binary image; (4) Contour detection;  (5) Independent zone detection; (6) Mesh generation.    }
  \label{fig:MEB_to_FEM}
\end{figure*}

In this work,  we focus  on a cluster of dendritic structures as shown in Fig.~\ref{fig:MEB_to_FEM}(1) which is extracted from Fig.~\ref{fig:MEB_image}. It is interesting to focus on this representative zone: its global shape  seems to be  comparable to a circular inclusion, while its internal tree-like structure may induce different acoustic features. To this aim, a non-dimensional analysis is expected to be carried out by scaling the above cluster of dendrite and the circular inclusion to a quasi-equivalent dimension. As shown in Fig.~\ref{fig:MEB_to_FEM}, this cluster of dendrite is encapsulated inside a square block. The square block containing a dendritic inclusion is then  used as elementary brick in the finite element simulation. 

Referring to the labels in Fig.\ref{fig:MEB_to_FEM}, the imaging procedure from the SEM image to a finite element mesh is: (1) The region of interest (ROI) is selected and extracted from the original SEM image. (2) Using  Matlab toobox Image Laberler, a pixel-level labeling is manually done in which pixels belonging to either the dendritc inclusion or to the matrix are labeled accordingly. 
Of course there are human factors in the steps of determining the two areas, but we have found that the manual method provides a better result than the tested automatic imaging methods like Canny edge detector \cite{Canny1986}, level sets\cite{RonaldFedkiw2002}, region growing \cite{Pal1993}, watershed\cite{Cousty2009}, etc.
 (3) The binary labeling information is then transformed into a binary image in which redundant parts are removed. The ROI  is the white zone and the matrix  is the black zone. (4) Contour detection gives accurate interfaces between the dendritic cluster and the matrix, and the width of the interfaces is one pixel. (5) The interfaces are segmented into separate zones, and each zone consists of a closed curve. Here,  we have 17 independent zones.   (6) To form the  final geometry, we sequentially import pixels coordinates of the 17 zones into COMSOL Multiphysics to create interpolation curves. Each closed curve creates a part of the dendritic inclusion inside which mechanical properties are homogeneous. The surrounding zone forms the matrix whose mechanical properties are different from the inclusion. Finaly,  P-1 triangle elements are employed to generate the displayed elementary mesh including both the inclusion and the matrix. It is essential that the number of nodes on the four boundaries of the square are  defined \textit{a priori} as the same and the nodes are equally spaced, for the reason of compatibility, given that we will  copy and arrange this elementary mesh in the horizontal direction and implement the periodic boundary condition on the upper and  lower boundaries as shown in Fig~~\ref{fig:geometry}.

\subsection{Volume fraction of the inclusion}
In this work, we prepared two models: one with dendritic inclusions, another with circular inclusions used for comparison. Since we are interested in the role of inclusion shape, we must ensure the best possible equivalence of the circular and dendritic inclusions, apart from the interface shape. 
As discussed previously, all the parameters involved in the material constitutive laws (the elastic moduli here) are scale-invariant, resulting in the same invariance for the whole model. We thus give all lengths in terms of the trivial length $L$, the distance between the inclusions. 
Equivalently, the frequency unit is chosen as $\omega_0=2\pi c_L/L$  which is used to define the unit time $t_0=L/c_L$ in the equation of motion. The relative size of the inclusions is then chosen in order to reproduce the data already obtained in Ref.\cite{Luo2019}  for circular inclusions. As in Ref. \cite{Luo2019}, the apparent diameter of the inclusion is thus chosen here as 5/6 L. It corresponds to the exact diameter of the circular inclusions, and to the longest axis of the dendritic inclusion. Therefore, the outer contour length is comparable between the two types of inclusions, which can be considered as the primary interface. However, the volume fraction for the dendritic inclusion (shown in Fig.\ref{fig:MEB_to_FEM}) and in 2D corresponding to a surface fraction is then measured as 28,35\%, while for the circular inclusion it is 54,54\%. The latter, which is close to twice the inclusion area of the former, intuitively allows for a more efficient scattering according to the results in Ref.\cite{Luo2019}. The geometrical and material parameters used are summarized in the Tab.\ref{table_volume_fraction}, and materials parameters are in Tab.\ref{table_para}. 

In the following, we will analyze the role of inclusion shape on acoustic attenuation in periodically arranged nanocomposites. As we are primarily interested to thermal transport, it is possible to scale the dendritic diameter to the nanometric lengthscale, pertinent for thermal transport at room temperature. For $L = 6 ~nm$ for example, as in Ref.\cite{Luo2019}, the reference frequency will be $\omega_0 = 8.34 $ THz, while for $L = 30 ~\mu m$, $\omega_0 = 1.67 $ GHz.

 \begin{table}[htbp]
\caption{List of inclusion (i) and matrix (m) volume fraction and largest axis for circular and dendritic inclusions}
\centering
\begin{tabular}{ccc} \hline
& Circular & Dendritic\\ \hline
$\Phi_i (\%)$ & 54.54& 28.35 \\ \hline 
$\Phi_m (\%)$ & 45.46& 71.65\\ \hline 
Largest axis & $\frac{5}{6}L$ & $\frac{5}{6}L$\\ \hline 
\end{tabular}
\label{table_volume_fraction}
\end{table}

\section{Numerical tools}

We used finite element numerical calculations to study the vibrational properties of a 2D semi-infinite elastic system with dendritic and circular inclusions positioned along a cubic lattice. The computational model consists of 9 squares, aligned in the horizontal direction. There is no initial inclusion-free block in this model. 
 The size of each square is defined as $L$,  thus determining the distance between inclusions as $L/6$. The wave-packet is generated imposing a displacement on the left side of the first square around $t=0$:
\begin{equation}
U(\omega,t)=U_0\exp(-\frac{(t-3t_0)^2}{2t_0^2})\sin(\omega t)
\label{eq:wavepacket}
\end{equation}
where $U_0$ is a constant value, $\omega$ is the frequency of this quasi-monochromatic excitation, and $t_0=\frac{3\pi}{\omega}$ is the half period of the excitation. A displacement parallel to the boundary corresponds to a transverse excitation, while the one perpendicular to the boundary to a longitudinal excitation. For the sake of simplicity, we will consider here only longitudinal excitations.

\begin{figure}[htbp]
\centering
\includegraphics[width=8cm]{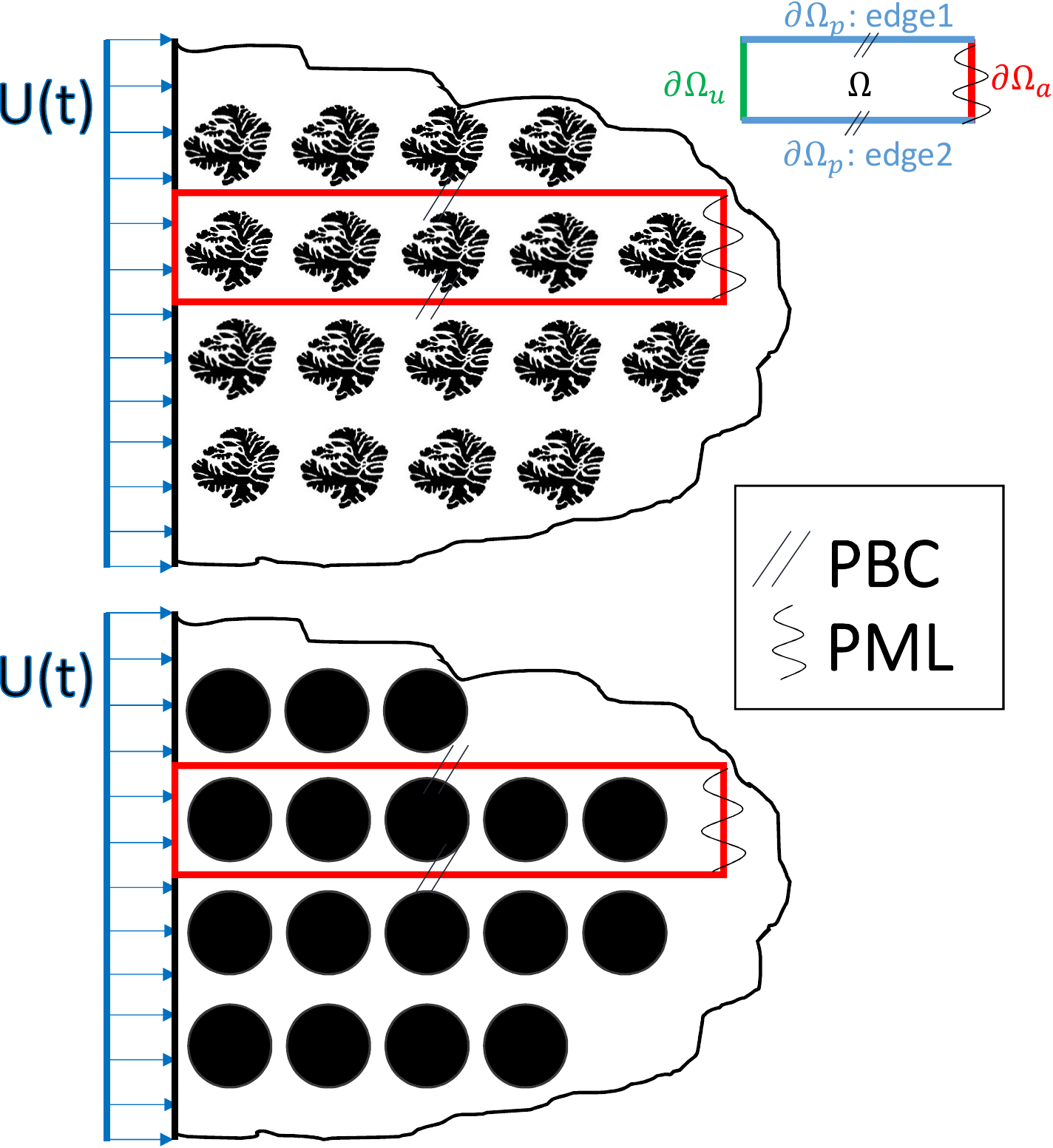}
  \caption{2D simulation model of a solid with dendritic (top panel) and circular (bottom panel) inclusions: this semi-infinite solid can
be represented by only modeling only the part inside the red rectangule with Periodic Boundary
Conditions (PBC) and Perfect Matched Layers (PML) as drawn; Black patterns represent the inclusions.
($\Omega$ represents simulation domain, $\partial \Omega$ indicates boundary conditions)}
  \label{fig:geometry}
\end{figure}

 As shown in  Fig.~\ref{fig:geometry},  periodic boundary conditions (PBCs) are implemented along the vertical direction at the top and bottom of the sample. Perfect Matched Layers (PMLs) are applied  on the right side to limit as much as possible waves reflection.  The technical details about the boundary conditions and the time integration scheme can be found in appendix in Ref.~\cite{Luo2019}.

Like in our previous works performed on a medium with periodic circular inclusions~\cite{Fusco2010}, the~matrix material is  linearly elastic with isotropic homogeneous elastic behavior characterized by a typical Young's modulus $E_m= 92.25$ GPa, the~mass density $\rho= 2303$~kg/m$^{3}$ and the Poisson ratio $\nu=0.347$ \cite{Beltukov2016}. For~the inclusions, the~Poisson's ratio is supposed to be the same, while the Young's modulus $E_i$ is taken as another control variable and defined as $E_i=E_m \times \frac{E_i}{E_m}$, this latter being the stiffness ratio 0.2 or 10. 
{Table~\ref{table_para} summarizes the values of the parameters used in this work:}

\begin{table}[htbp]
\caption{List of Parameters and Reference~dimensions}
\centering
\begin{tabular}{cccc} \hline
{$E_m$} (GPa) & {$\nu_m$} & $E_i/E_m$ & $\nu_i$ \\ \hline
92.25  ~\cite{Fusco2010} & 0.347 ~\cite{Fusco2010} & 0.2 or 10  & 0.347 \\ \hline \hline
 $\rho$ (\si{kg/m^{3}}) & $c_L$ (\si{m/s})& $\omega_0$ (\si{rad/s} )&  $\omega/ \omega_0$  \\ \hline
2303  ~\cite{Fusco2010} & $7966$ & $8.34 \times 10^{12}$ &  0.3-4.8  \\ \hline
\end{tabular}
\label{table_para}
\end{table}

\section{Acoustic transport in an isotropic homogeneous material with dendritic inclusions}

A set of transient simulations of longitudinal wave-packet propagation is done using FEM for both the medium with dendritic and circular inclusions. From the results, we analyze the envelope of the kinetic energy in order to identify the attenuation regime. In addition, the penetration length and diffusivity are calculated to compare the attenuation ability for the two types of inclusions. Finally, long-wavelength and instantaneous sound speeds are estimated for some representative cases.
In the following,   the kinetic energy $E_k$ is normalized by the maximum value at $x=0$.  

\subsection{Envelope of the kinetic energy}

As said before, the~wave-packet is created by imposing a displacement on the left side of the sample. Its propagation is then followed along the sample, in~the $x$ direction.
Due to the presence of interfaces, and~related spatial inhomogeneities, the~wave-packet wave-vector $\vec k$ does not remain constant, the~wave-packet being scattered by the inclusions. To~understand how such scattering affects the energy transfer, we measure the envelope of the kinetic energy induced in the system by the propagation of the wave-packet, summed over the y-direction. The~energy envelope is defined for each excitation frequency $\omega$ as
\begin{equation}
P_{\omega}(x)=\max_t E_k(x,t)
\label{eq:envelope}
\end{equation}
where $E_k(x,t)$ is the instantaneous kinetic energy supported by the frame located in $x$ with width $\Delta x=L/60$.

\begin{figure*}[htbp!]
\centering
\includegraphics[width=18cm]{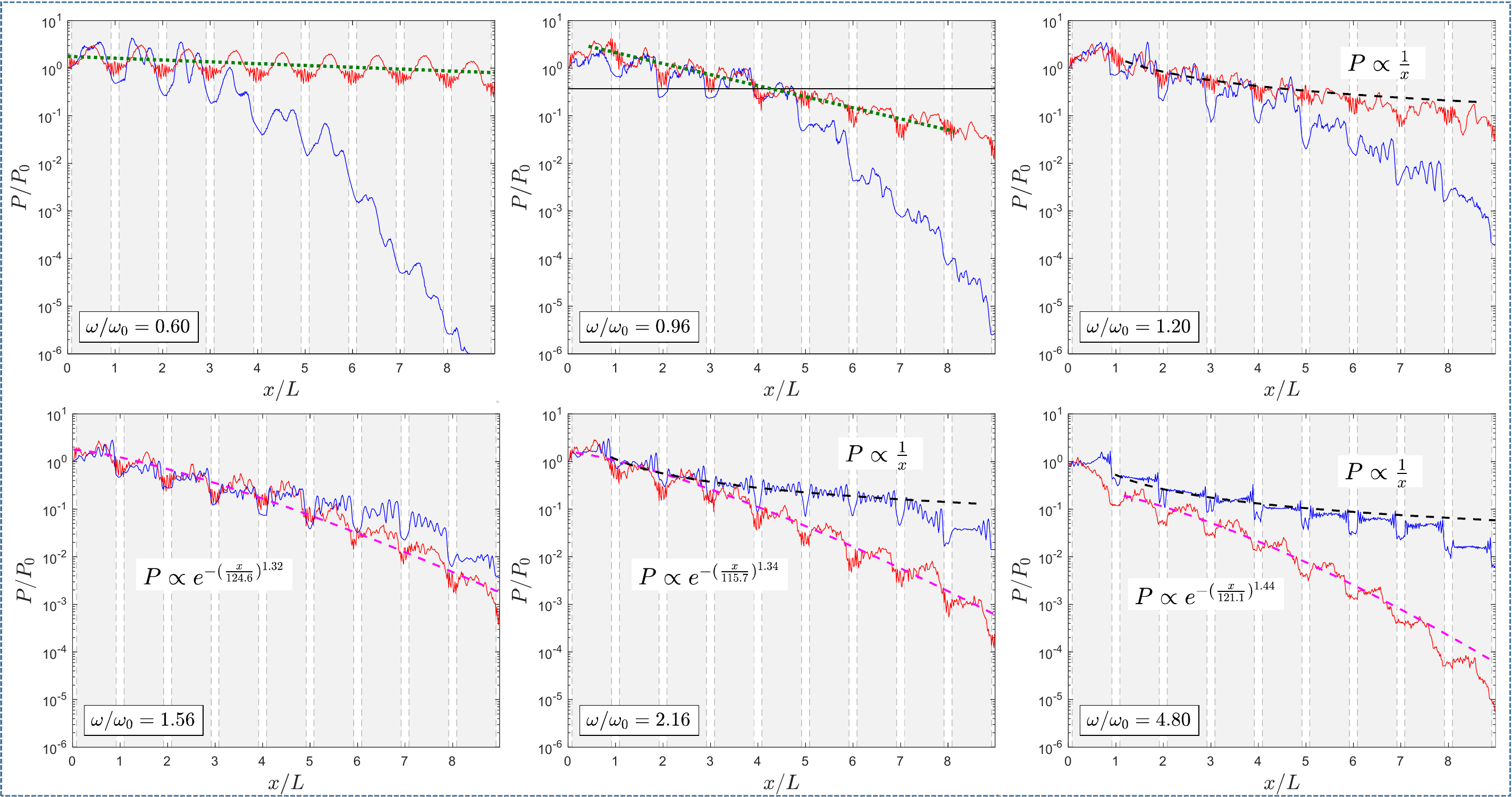}
  \caption{Comparisons of envelopes between circular (blue) and dendritic (red) inclusions with $\frac{E_i}{E_m}=0.2$  for different normalized frequencies $\omega/\omega_0$  (where $\omega_0 = 2\pi c_L / L  $) on a semi-log scale: total simulation time  $8400~dt$ with $dt = \Delta l / c_L $ ($\Delta l$ is linked to mesh size). For the case  $\omega/\omega_0$=0.60 and 0.96, green dotted line represents an exponential decay. For the case $\omega/\omega_0$=1.20, 1.56, 2.16 and 4.80, the black dashed and violet dashed lines represent $1/x$ and compressed exponential fit, respectively. For the case $\omega/\omega_0$ = 0.96, as an example, we report  the level $P=P_0/e$, with $P_0$ the starting energy value as a black solid line: its intersection with the envelop  determines the penetration length.}
  \label{fig:dend_comparisons_envelope}
\end{figure*}

\begin{figure*}[htbp!]
\centering
\includegraphics[width=18cm]{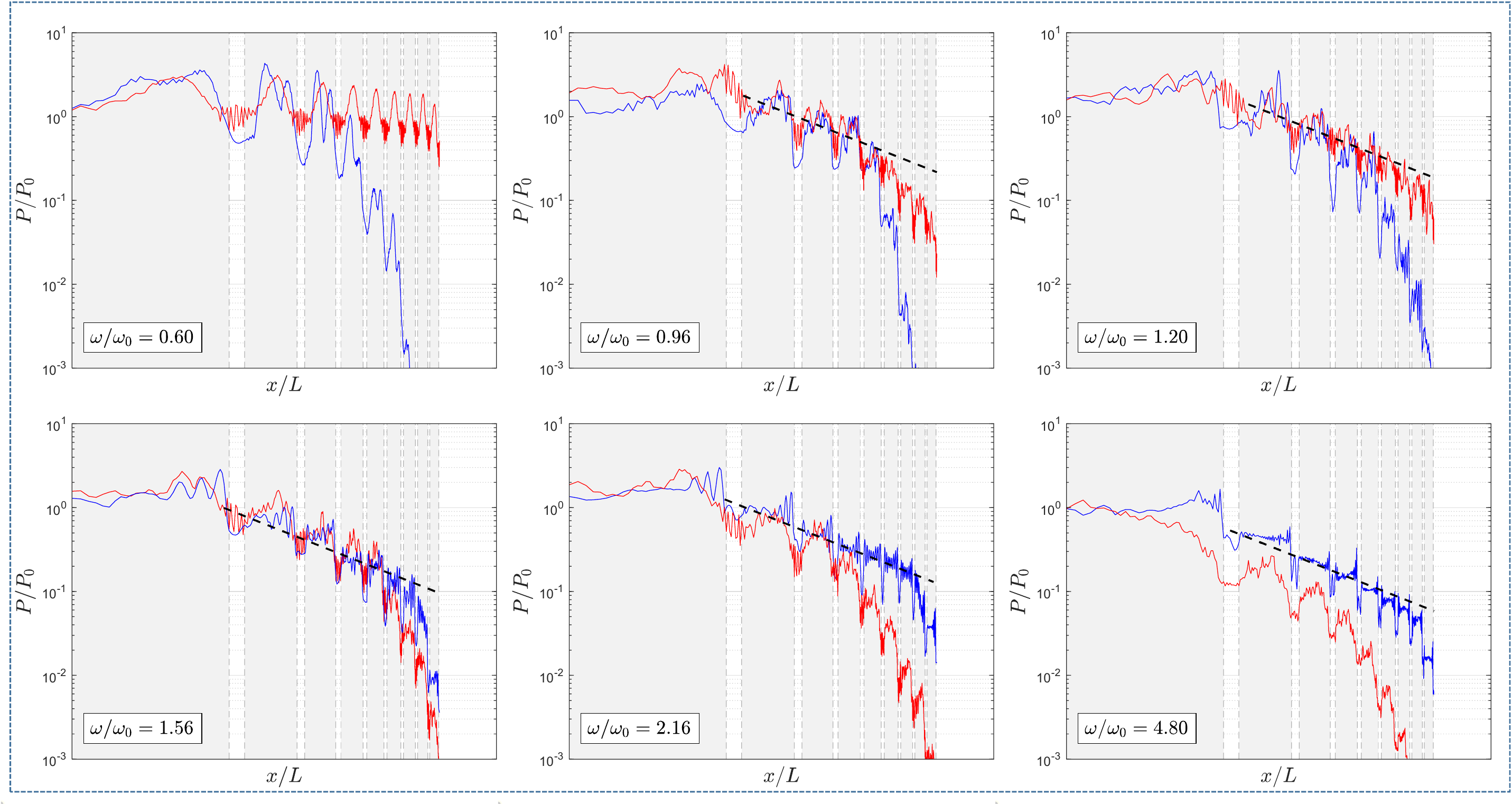}
\caption{ Fig.\ref{fig:dend_comparisons_envelope} on a   log-log scale. Blue  lines are for the circular inclusions and  red lines  for the dendritic inclusions.  In this representation the diffusive $A/x$ behavior, with $A$ a fitting parameter, corresponds to a linear decrease, as evidenced by the black dashed lines. }
  \label{fig:dend_comparisons_envelope_loglog}
\end{figure*}

In Fig.\ref{fig:dend_comparisons_envelope}, we present the envelopes of the kinetic energy as a function of position on a semi-log scale and for different normalized frequencies $\omega /  \omega_0$ with  $\frac{E_i}{E_m}=0.2$.  
Surprisingly, for low frequencies, up to $\frac{\omega}{\omega_0}=1.2$ included, the circular inclusions strongly attenuate the wave-packet, which is almost unaffected by the presence of the dendritic inclusions. The strong effect of circular inclusions has been previously understood as the consequence of a localization of the energy due to a resonance of the inclusions that keeps the acoustic energy \cite{Luo2019}. However, the irregularly shaped dendritic inclusion does not lead to such resonance phenomena. As such, we can conclude that for low frequencies, long wavelengths, the high interface density of the dendritic inclusion is not dominant in determining the wave packet attenuation.
The situation is reversed above $\frac{\omega}{\omega_0}$=1.56. 
From this frequency,  the dendritic inclusions become more efficient in attenuating the wave-packet and such attenuation strongly increases with increasing frequency, \textit{i.e.} decreasing wavelength. In the circular case,  the attenuation is almost constant for frequencies above $\frac{\omega}{\omega_0}$=2.16, suggesting a saturation of the attenuation effect of circular inclusions in the matrix. We can understand the change of attenuation efficiency regime as due to the major importance of the tree-like interface in dendritic inclusions at wavelengths comparable with the dendritic structure lengthscale, \textit{i.e.} of order of $ \approx L/10$. 

We can also identify the different regimes directly from the envelopes of the kinetic energy.
Indeed, in case of a weak scattering (propagative regime) a global exponential attenuation similar to a Beer-Lambert law can be observed~\cite{Swinehart1962,Beltukov2018,Luo2019}
\begin{equation}\label{eq:BL}
P_{\omega}(x)\propto \exp(-x/{\Lambda(\omega)})
\end{equation}
while for strong scattering, for example due to the larger rigidity contrasts as in this work ($E_i/E_m= 0.2$ or 10), the~algebraic attenuation of the envelope
\begin{equation}
P_{\omega}(x)\propto 1/x
\end{equation}
is the signature of a diffusive process (diffusive regime)~\cite{Beltukov2018,Luo2019}.   Since Fig.\ref{fig:dend_comparisons_envelope} is given on a semi-log scale, the propagative regime which follows an exponential decay gives a straight line with a negative slope. In Fig.\ref{fig:dend_comparisons_envelope}, as exemplified by the green dotted line for $\omega/\omega_0$=0.60 and 0.96 in the case of dendritic inclusions,  an evident exponential decay  can be recognized at low frequencies. At higher frequencies,  we find $1/x$ behavior  for both the case of dendritic ($\omega/\omega_0=1.20$) and circular inclusions ($\omega/\omega_0=2.16$ and $4.80$), which can be then verified on the log-log plot where the $1/x$ behavior corresponds to a straight line with a fixed slope, as shown in Fig.\ref{fig:dend_comparisons_envelope_loglog}. Interestingly, for dendritic case with $\omega/\omega_0=1.56, 2.16$ and $4.80$, the envelope  follows neither the B-L fit nor the diffusive fit. If we use a compressed exponential function, such as $P \propto \exp{(-(x/\Lambda)^\beta)}$ with $\beta > 1$, the best fit value of  $\beta$ is 1.32, 1.34 and 1.44 respectively. 
Such a compressed exponential behavior   marks a strong difference from the case of circular inclusions, where the diffusive regime is well followed, with a reduction of the oscillations amplitude, as better visible in Fig.\ref{fig:dend_comparisons_envelope_loglog}. In the case of dendritic inclusions, instead the novel compressed exponential regime exemplifies in fact the succession of two regimes: if at the beginning the diffusive law seems to be followed, this leaves space very rapidly to a much stronger attenuation. In both regimes oscillations
are present, similarly to the circular inclusions
case, with a possibly larger amplitude at high frequency, indicating that the energy is
pinned within the inclusions.

\begin{figure*}[htbp]
\includegraphics[width=18cm]{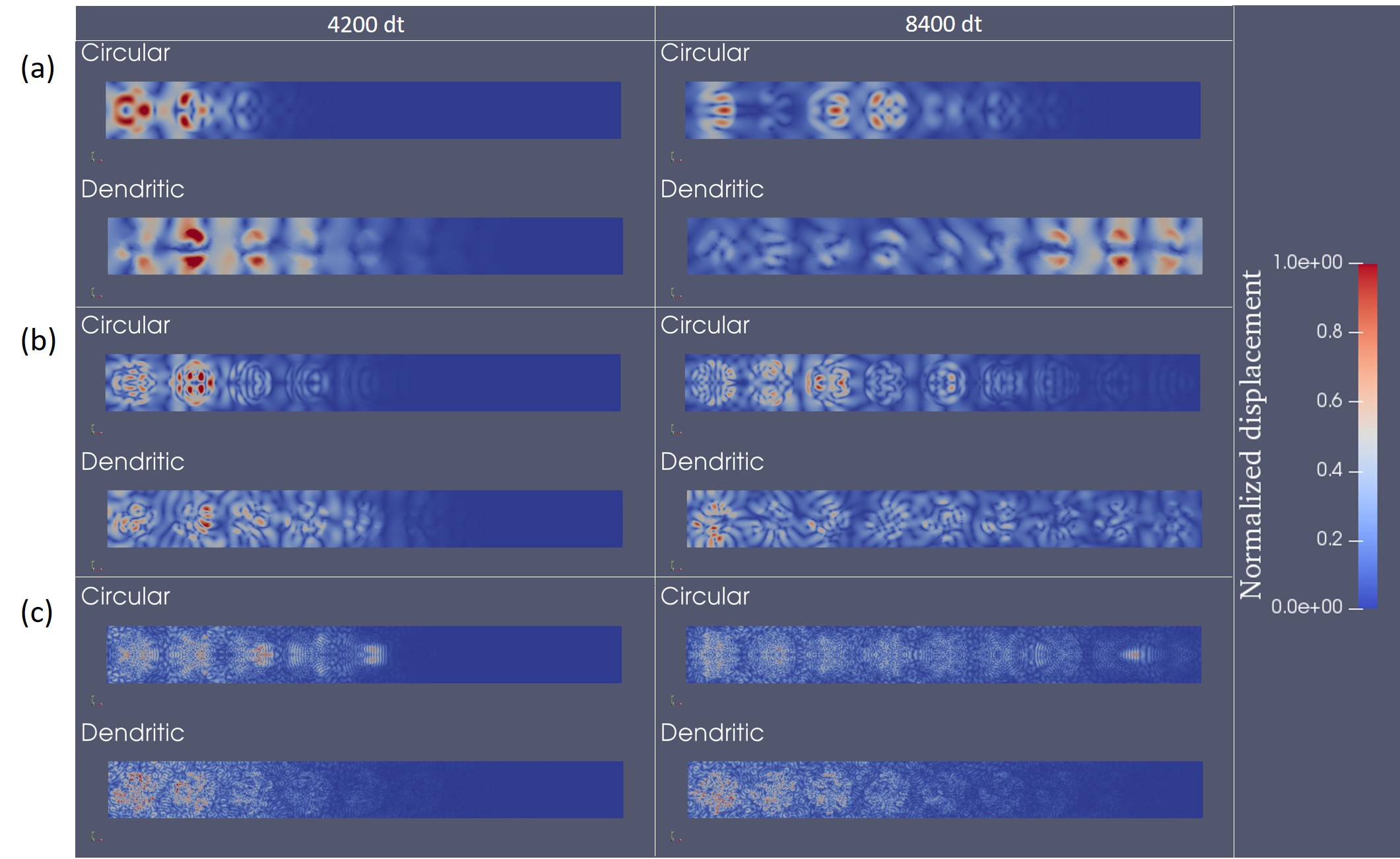}
  \caption{Snapshots of 3 case at 4200 $\times$ dt and 8400 $\times$ dt with $\frac{E_i}{E_m}=0.2$ for  circular and dendritic inclusion: (a)  $\omega / \omega_0=0.6$ (b) $\omega / \omega_0=1.2$ (c) $\omega / \omega_0=4.8$. (see Supplementary Materials for the corresponding videos ).}
  \label{fig:dend_snapshot}
\end{figure*}

In order to get insight into the appearance of this new attenuation regime in the dendritic sample, we report in 
 Fig.\ref{fig:dend_snapshot}, snapshots of the displacement field  for low frequency ($\omega/\omega_0 = 0.6$), medium frequency ($\omega/\omega_0 =1.2$) and high frequency ($\omega/\omega_0 =4.8$) at  half time (4200 dt) and at the final time (8400 dt)  for both samples. For $\omega/\omega_0 = 0.6$, the energy is localized and pinned to the first inclusions in the case of circular inclusion but spreads rapidly  in the dendritic ones without huge dispersion meaning that the wavefront propagates ahead followed by an energy tail.  At the medium frequency $\omega / \omega_0 = 1.2$, the energy is dispersed in space and the attenuation length in the two samples is quite comparable. At the high frequency $\omega / \omega_0 = 4.8$, the energy is scattered violently at the inclusion-matrix interfaces. In the case of circular inclusions, energy is only scattered few times when crossing the circular interface. In addition, due to the large curvature of the circle compared with the short wavelength, there is little but existing energy which  keeps spreading ahead along the x direction.
On the contrary, for dendritic inclusion, energy is totally scattered due to the  random orientations of the normals at the interface and high interface density.

To conclude, different regimes can be identified: (1) exponential attenuation is observed mainly in the low frequency range, which relates directly to the propagative contribution; (2) diffusive regime is observed at higher frequencies, which can be identified through the $P \propto 1/x$ behavior and is related directly to the diffusive contribution; (3) localized (mixed diffusive-localized) regime can be found at some specific frequencies, for example $\omega/\omega_0=0.6$ for circular inclusions, which most effectively prevent the energy transport; (4) compressed exponential attenuation  regime is found in the high frequency range for dendritic inclusions, where the initially diffusive regime seems to be replaced by a stronger attenuation, leading to a global compressed exponential behavior. This results from  the combined effect of periodicity and complex interface shape of the dendritic structure.  

\subsection{Penetration length \& diffusivity}

In the previous section, by comparing the kinetic energy envelopes, we have given a first picture of the attenuation ability   in the solids with two types of soft inclusions. 
Now, we will quantify this attenuation by calculating the penetration length and the diffusivity as a function of the frequency in two extreme cases of rigidity contrast, being $\frac{E_i}{E_m}$ = 0.2 and 10.0.

First,  we look at the long-time penetration length, defined as the traveled length above which the energy per unit length remains always smaller than the maximum excitation energy per unit length divided by $e$.  An example is reported in Fig.\ref{fig:dend_comparisons_envelope} for $\omega/\omega_0=0.96$, where the abscissa of the intersection between the envelope and the $A/e$, with $A$ the maximum excitation energy  (black solid line) determines the attenuation length.

\begin{figure}[htbp]
\centering
\includegraphics[width=8cm]{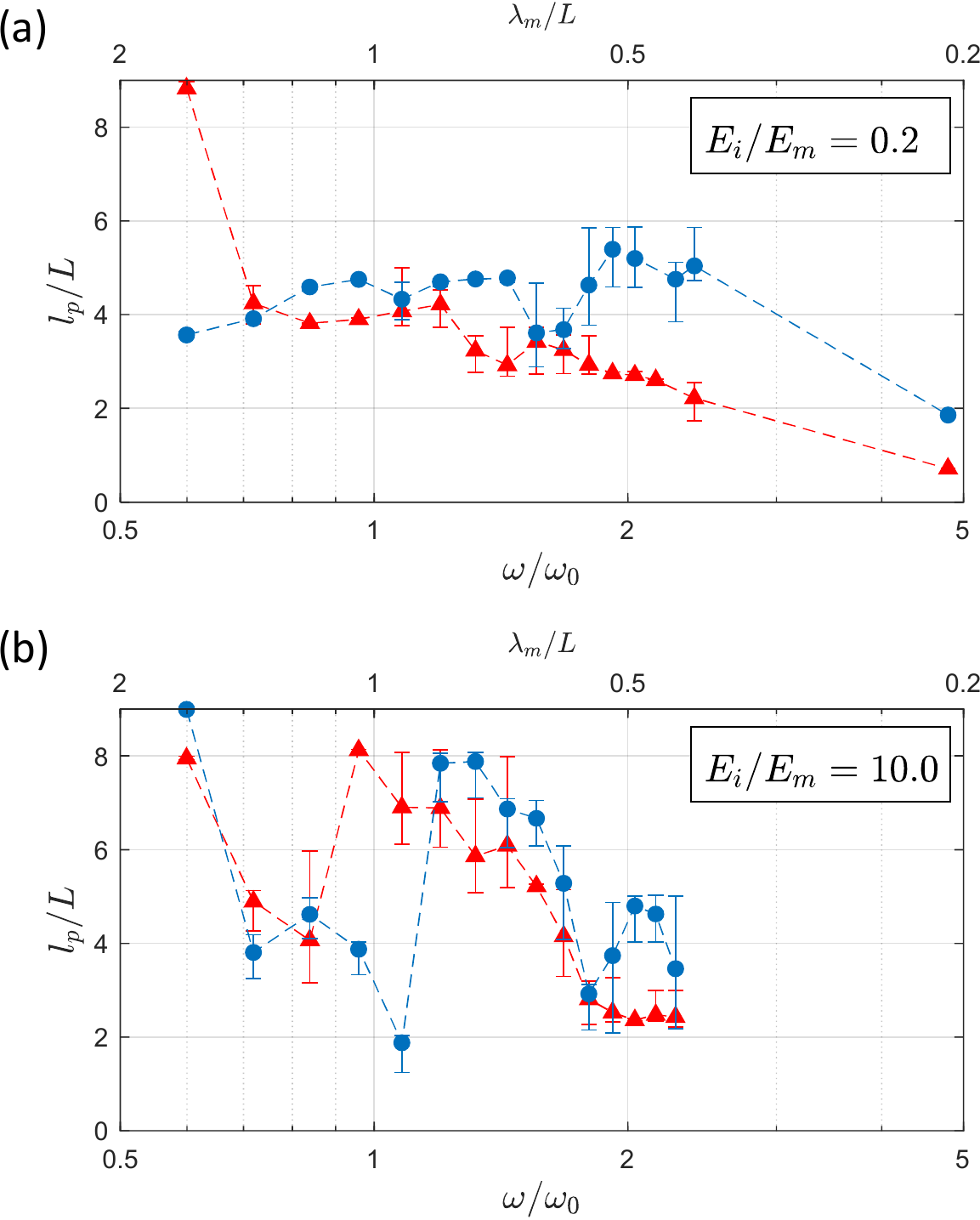}
  \caption{Normalized penetration length of longitudinal wavepackets in samples with circular (blue circles) and dendritic (red triangles) inclusions for  $E_i/E_m=0.2$  (panel a) and $10.0$ (panel b). The corresponding wavelength scale is reported on the top axis.}
  \label{fig:dend_comparisons_pene_length}
\end{figure}

The results  of  the normalized penetration length  $l_p/L$  are shown in Fig.\ref{fig:dend_comparisons_pene_length}  for both dendritic  (red triangles) and circular inclusions (blue circles). In both cases of  $\frac{E_i}{E_m}$, at low frequencies, the dendritic inclusions do not exhibit their high-interface-density advantage. Still, as wavelength decreases, dendrite shape begins showing a better performance to attenuate energy transfer with the reduced crossover frequency $\frac{\omega_c}{\omega_0}$ ranging between 1 and 1.5 depending on the stiffness ratio.   Generally, for $\omega > \omega_c$, values of the penetration length is systematically shorter in the medium with dendritic inclusions, which indicate a high potential for such samples for reducing the propagative contribution to thermal transport. Interestingly, except for the soft inclusions case at the highest frequency  ($\omega/\omega_0=4.8$), almost all the penetration lengths are greater than the characteristic length $L$.  Finally, it is worth noticing that for $E_i/E_m$ = 10, a local minimum appears in the penetration length for both types
of inclusions, being around $\omega/\omega_0$=1 in the circular case and $\omega/\omega_0$=0.8 in the dendritic case. This effect has already been reported and ascribed to the collective resonance of the nanoparticles 
in the low frequency range, and related to the effective acoustic
impedance of the composite, its frequency position being proportional to $\sqrt{Ei/Em}$ \cite{Luo2019}. The slight dependence of this position on the inclusion shape is related to the effective stiffness of the medium. Indeed, the ratio of the two positions results 1.25, which is very close to the ratio between the square roots of the effective stiffnesses ($E_{eff}$) of the two media, that we calculate in the next section and  can be found in Tab.\ref{tab:effective_properties_ch3_dendrite}. This suggests the existence of a mixing law of the effective mechanical properties for the bi-phase composites.

In order to investigate the effect on the diffusive contribution to thermal transport, we
first identified the situation where the diffusive regime can be identified from the time evolution of the wave-packet average position $\langle x \rangle (t)$. To this aim, we have calculated this positions as:
\begin{flalign}
\langle x \rangle(t)=\frac{\sum_i x_i E_k(i,t)}{\sum_i E_k(i,t)}
\label{eq:stas1}
\end{flalign}
where $x_i$ is the position of the $i^{th}$ frame with width $\Delta x$ in the x-direction, and~$E_k(i,t)$ is the instantaneous total kinetic energy supported by that frame. 
In the case of a diffusive process, the squared deviation $\sigma^2(t)$ is proportional to the time $t$, with a slope related to the one-dimensional diffusivity
\begin{equation}\label{eq:diff}
\sigma^2(t)=2D(\omega)t.
\end{equation}
where $D$ is the diffusivity and  the spreading $\sigma$ reads

\begin{flalign}
\begin{split}
\sigma(x,t)&=\sqrt{{\langle \left(x -\langle x \rangle \right)^2 \rangle}}\\
&=\sqrt{\frac{\sum_i (E_k(i,t)\times x_i^2)}{\sum_i E_k(i,t)}-{\langle x \rangle}^2}
\label{eq:stas2}
\end{split}
\end{flalign}

\begin{figure}[htbp]
\includegraphics[width=8cm]{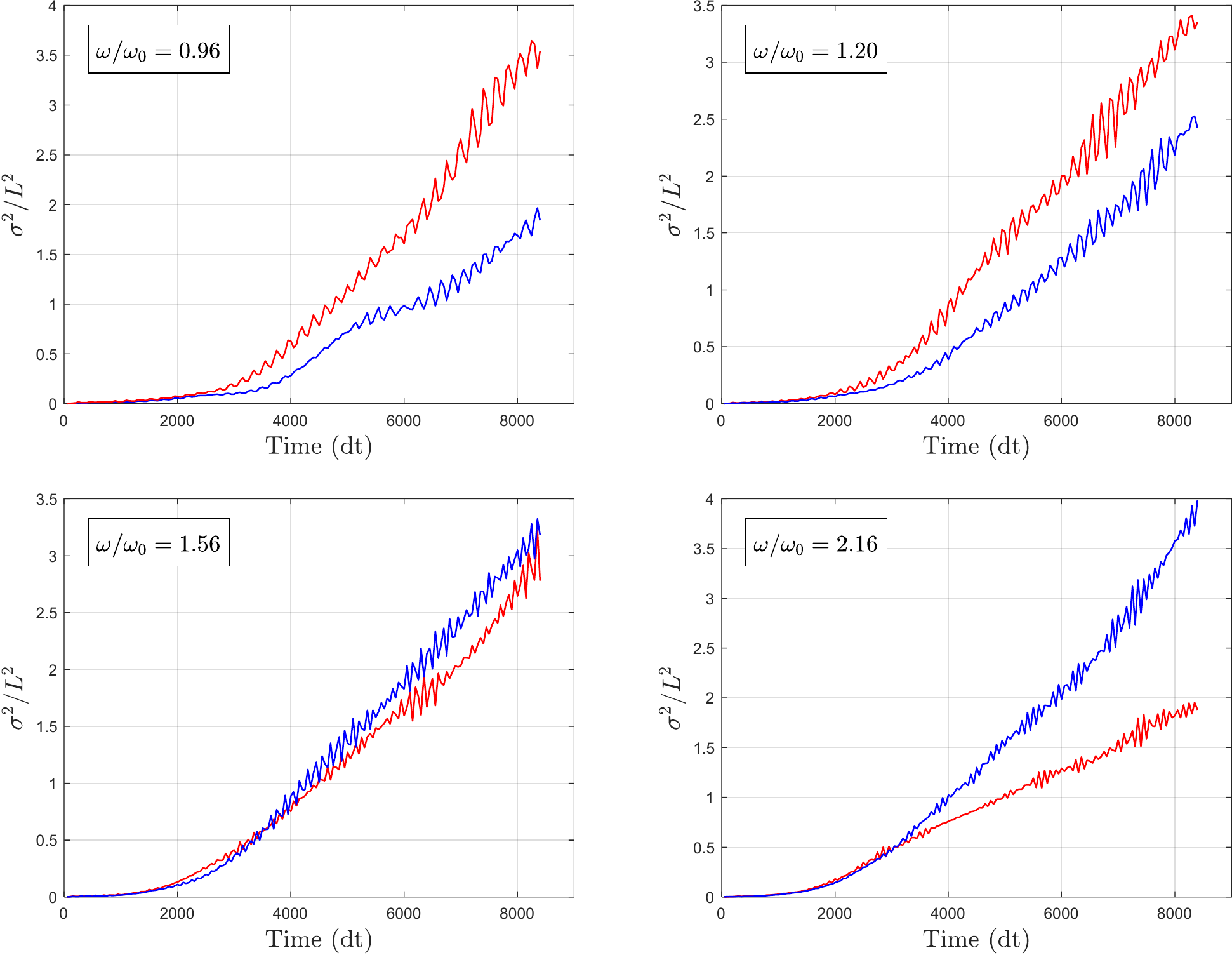}
  \caption{Comparison of $\sigma^2/L^2$ of random longitudinal wavepackets in samples with circular (blue) and dendritic (red) inclusions with $E_i/E_m=0.2$ for different normalized $\omega/\omega_0$}
  \label{fig:etalement}
\end{figure}

As an example,  $\sigma^2$  of random longitudinal wavepackets normalized by $L^2$ as a function of time step is  shown in Fig.\ref{fig:etalement} for four  frequencies. In the high frequency range, in both samples we clearly find a  $\sigma^2 \propto t$  relation, indicating the diffusive transport of energy,  from which the diffusivity $D$ can be fitted. 
It is worth underlining that in the sample with dendritic inclusions we find a diffusive time evolution of the wave-packet even at frequencies for which the anomalous stronger-than-diffusive attenuation appears. 
Concerning the sample with circular inclusions, interestingly we observe the presence of a short plateau for $\omega/\omega_0=0.96$, which can be understood as the signature of localization.
Indeed, in the localized regime, the wave-packet is pinned so that $\langle x \rangle=cste$ as well as  $\sigma^2$, giving thus a plateau.

\begin{figure}[htbp]
\centering
\includegraphics[width=8cm]{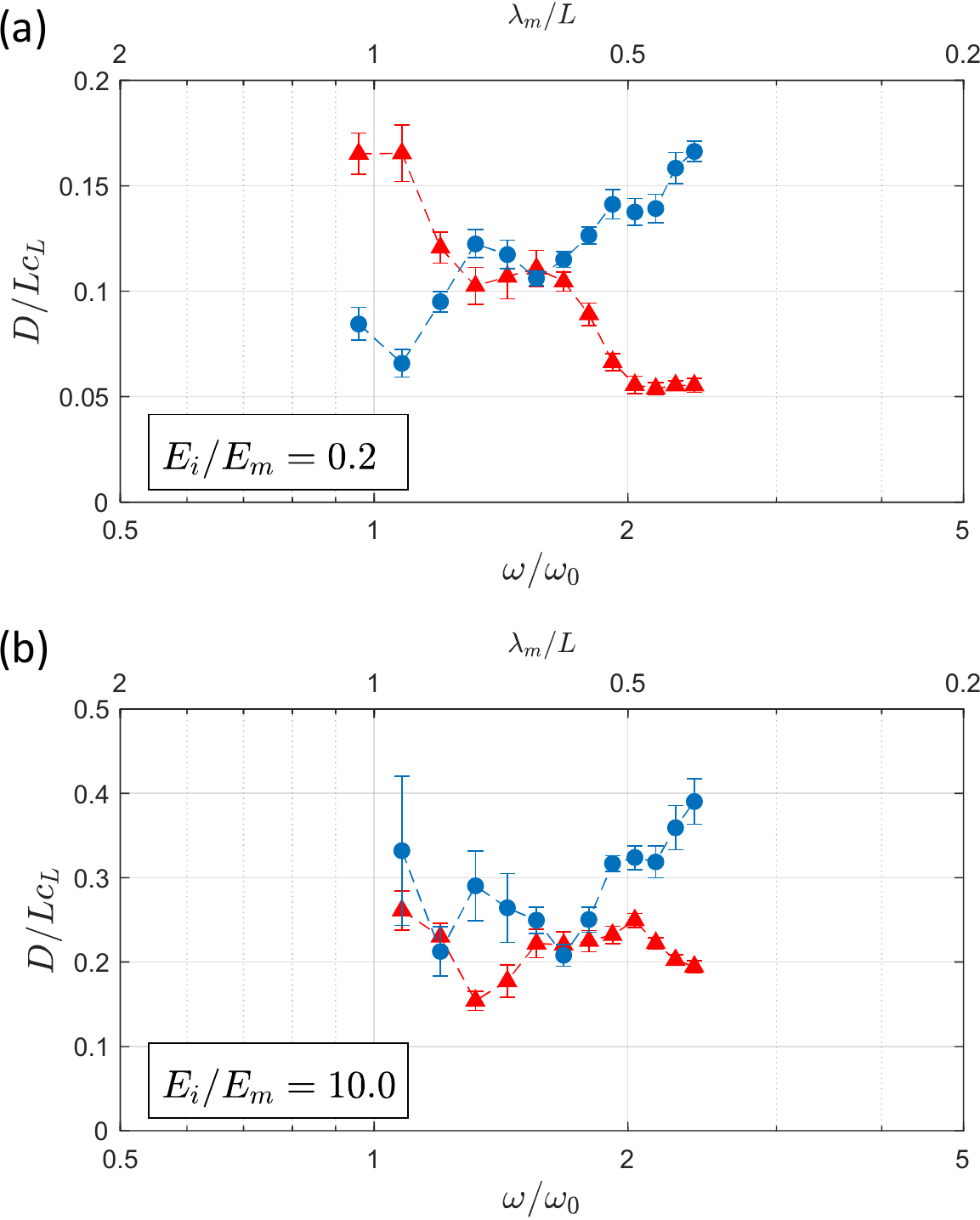}
  \caption{  Normalized diffusivity of random longitudinal wavepackets in samples with circular (blue circles) and dendritic (red triangles) inclusions for $E_i/E_m=0.2$  (panel a) and $10.0$ (panel b).  The corresponding wavelength scale is reported on the top axis. }
  \label{fig:dend_comparisons_diffusivity}
\end{figure}

Results of diffusivity are shown in Fig.\ref{fig:dend_comparisons_diffusivity}. 
We can draw similar conclusions as for the penetration length, that is, at high frequencies,  the dendritic shape of the  inclusions lead to a lower diffusivity than in the circular case.
If this conclusion seems to be universal, independently on the stiffness ratio, it is interesting to note that for $\frac{E_i}{E_m} = 10$ the diffusive contribution is reduced by the dendritic shape of the inclusions at all wavelengths. 
If  now we compare the diffusivity between $\frac{E_i}{E_m} = 0.2$ and $\frac{E_i}{E_m} = 10$, we remark that the overall diffusivity in the soft inclusions case is lower than that of the hard inclusions case, which is coherent with the conclusions in Ref.\cite{Luo2019}.

\subsection{Sound velocity}
In composites nano-phononic materials, the wave propagation velocity will be mainly influenced by two factors: the effective rigidity of the medium and the properties of the  interfaces.  In the following, we will investigate two types of wave speed : 1) effective wave speed at low frequencies and 2) the instantaneous wave speed. The latter is especially useful when it is difficult to get a stationary speed (independent of time), due to the strong elastic heterogeneity of the medium.

\subsubsection*{Long-wavelength speed}
Long-wavelength speed for longitudinal waves can be calculated as:
\begin{equation}
    c_{L,eff}=\sqrt{\frac{E_{eff}(1-\nu)}{\rho(1+\nu)(1-2\nu)}}
\end{equation}
where $E_{eff}$ is the effective Young's modulus of the composite estimated by the Reuss model, which  states that  the elastic modulus of a composite can be expressed as:
\begin{equation}\label{eq:ch3_reuss_model}
E_{eff} = \frac{1}{\Phi_i \times \frac{1}{E_i} + \Phi_m \times \frac{1}{E_m}}
\end{equation}
where $E_i$ ($E_m$) is the Young's modulus of the inclusion (matrix) and $\Phi_i$ ($\Phi_m$ ) is the volume fraction of the inclusion (matrix).

   We summarize the material properties, the effective Young's modulus and the effective longitudinal wave speeds  in the Tab.\ref{tab:effective_properties_ch3_dendrite}. From the estimated wave speeds, it is clear that the long-wavelength wave speed increases compared to the homogeneous solid with $E_i/E_m=1$ in the case $E_i/E_m >1$ and decreases in the case  $E_i/E_m <1$ whatever the shape of inclusion is. 
We find that, both for a more rigid or a softer inclusion, the long wavelength speed is more strongly affected for a larger volume fraction, \textit{i.e.} in the case of the circular inclusions. The speed at such wavelengths is thus essentially determined by the elastic moduli of the phases and the volume fraction of the secondary phase, independently on the inclusion shape. 
   
This calculation does not reflect however the diversity of behaviors that we have seen in the previous sections, and specifically the fact that the dendritic shape can strongly affect both propagation and energy diffusion at some wavelengths and the wave-packet can be trapped between inclusions. Such differences will affect the wave-packet speed at wavelengths comparable with the nanostructure, \textit{i.e.} when the wave-packet is actually perturbed by the presence of the inclusions in its propagation behavior. For this reason, we address in the following the instantaneous speed.

\begin{table}
\caption{Material properties, the  effective Young's modulus and the effective longitudinal wave speed for composites with circular and dendritic inclusion with Poisson coefficient $\nu =0.347$ and density $\rho = 2303$ kg/$m^3$.} 
\centering
\begin{tabular}{|c|c|c|c|c|c|c|}
\hline
 & \multicolumn{3}{|c|}{\thead{Circular}} & \multicolumn{3}{|c|}{\thead{Dendritic}} \\ \hline
  $E_m$ (GPa)&\multicolumn{6}{|c|}{92.25} \\ \hline
 $E_i/E_m$ & 0.2& 1.0 &10  & 0.2& 1.0& 10\\ \hline
  $E_i$ (GPa)& 18.45& 92.25 & 922.5 & 18.45& 92.25& 922.5\\ \hline
   $\Phi_m$ (\%) &  \multicolumn{3}{|c|}{45.46}  & \multicolumn{3}{|c|}{71.65}  \\ \hline
   $\Phi_i$  (\%) &   \multicolumn{3}{|c|}{54.54}  & \multicolumn{3}{|c|}{28.35} \\ \hline 
   $E_{eff}$ (GPa)      & 28.99 & 92.25 & 181.19&43.23  & 92.25& 123.85\\ \hline
   $c_{L,eff}$ (m/s)    & 4466.1& 7966.1 &11164.3  & 5453.2 & 7966.1 & 9230.3\\ \hline
\end{tabular}   \label{tab:effective_properties_ch3_dendrite}
\end{table}

\subsubsection*{Instantaneous wave speed at high frequencies}
When the wavelength approaches the nanostructure lengthscale (size and inter-distance between scatterers), the phonon-interface scattering becomes more important and wave speed begins to deviate from the long-wavelength value. As the wave-packet moves in a highly heterogeneous medium, its velocity is not homogeneous in space nor in time, the wave-packet being scattered in different directions, backward included.
Therefore,   we need to calculate the \textit{instantaneous speed}  defined as:

\begin{equation}\label{eq:instantaneous_speed}
c_{ins} = \frac{\partial <x>(t)}{\partial t}
\end{equation}
where $<x>$ has been defined in Eq.\ref{eq:stas1} and needs to be smoothed, because the energy oscillates back and forth in the medium, 
due to the multiple reflections from the interfaces. This can be clearly seen in the Appendix B,  Fig.\ref{fig:dend_instantaneous_speed_1}. The real, non-smoothed, instantaneous speed will necessarily sharply fluctuate and even assume negative values for backreflections. By smoothing $<x>$, Eq.\ref{eq:instantaneous_speed}  will instead give us the average speed of energy transport in a relatively short time.   In this work, we use the Bezier interpolation as detailed in appendix B.

\begin{figure}[htbp!]
\centering
\includegraphics[width=8cm]{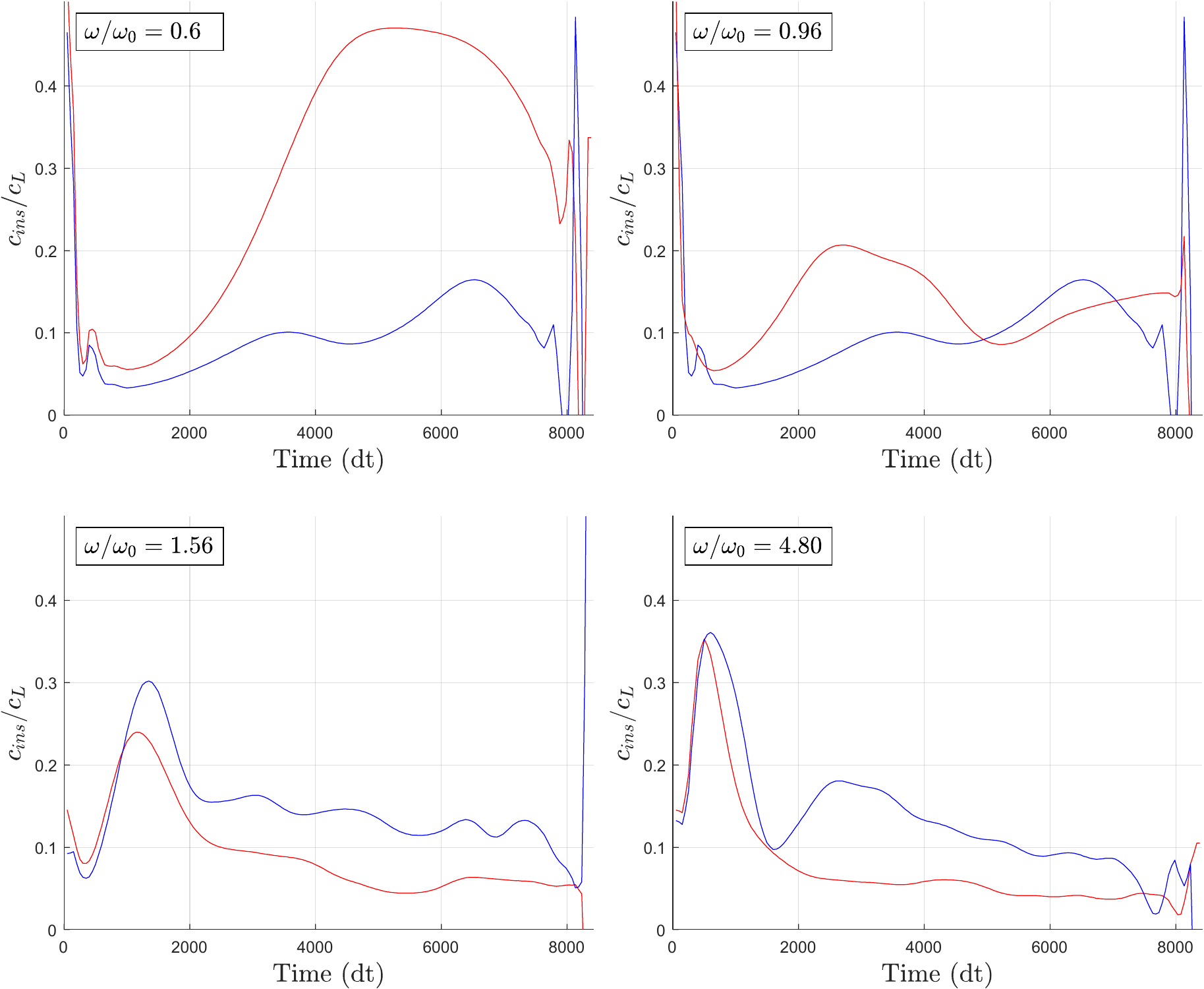}

  \caption{Instantaneous wave speed $c_{ins}$ normalized to $c_L$ in samples with dendritic (red) and circular (blue) inclusions and stiffness ratio $E_i/E_m=0.2$. See the text for the details on the calculation.}
  \label{fig:dend_instantaneous_speed_2}
\end{figure}

We select four  frequencies  from low to high: $\omega / \omega_0$ = 0.6, 0.96, 1.56 and 4.8  with $E_i/E_m = 0.2$. The corresponding instantaneous wave speeds are calculated and shown in Fig.\ref{fig:dend_instantaneous_speed_2}. 
Note the initial increase due to the establishment of the wave packet inside the sample, at earlier times for higher frequencies. Only the part after this initial increase must be considered in the following.
First, for $\omega / \omega_0$ = 0.6 and dendritic inclusions, $c_{ins}$ exhibits a plateau showing a quasi-constant instantaneous speed
$\approx 0.45 c_L = 3580 ~m/s$. This speed is always lower than the estimated long-wavelength value reported  in  Tab.\ref{tab:effective_properties_ch3_dendrite} which is $5453  ~m/s$ for $E_i/E_m=0.2$, indicating a slight attenuation of energy transport. 
 Conversely, this plateau  does not exist for the circular inclusions case because it is already in the diffusive-localized regime where energy is pinned in the first inclusions. The instantaneous velocity in the circular inclusions case is always lower than the one in the dendritic inclusions case, meaning that energy moves more slowly in the medium with circular inclusions at all time.
 This is in agreement with the smaller penetration length found at this frequency in the circular inclusions sample than in the dendritic one, as shown in Fig.\ref{fig:dend_comparisons_pene_length}(a). For  $\omega / \omega_0$  = 0.96, the instantaneous speeds in the two samples are quite similar to each other, which, once again, is in good agreement with our findings on the penetration length. 
 $\omega / \omega_0$  = 0.96 can be considered as a separation  point, because at the following two frequencies $\omega / \omega_0$  = 1.56 and 4.8, the instantaneous speed in the circular inclusions samples is systematically larger than in the dendritic case, meaning that at high frequency the dendritic sample is more efficient in slowing down and reducing energy transport.

To conclude, the analysis of the  instantaneous speed gives a more obvious picture on the crossover of the attenuation performance from the circular inclusions case at low frequencies to the dendritic inclusions case at high frequencies for $E_i/E_m$ = 0.2, especially when the  wave-vector  no longer exists. 
Interestingly, this velocity is not stationary in the frequency range studied here.

\section{Discussion and Conclusions}

To conclude, we have shown the effect of a complex shape in the periodic pattern of a 2D nano-phononic crystal. We have compared the sound attenuation performance between two shapes of  inclusions: circular and dendritic.  
Our results show that the multi-branching tree-like form of dendrites enhances  phonon-interface scattering and phonon attenuation specifically for wavelengths comparable with the dendritic  structure lengthscale regardless the rigidity ratio.
Unlike the circular inclusion which has only one characteristic length, the sub-interfaces inside the dendritic inclusion provide a continuous source of scattering leading to an increasing sound attenuation. This leads to a stronger reduction of both the penetration length and the apparent diffusivity in samples with  dendritic inclusions when the wavelength becomes smaller than the first  characteristic length even for far smaller volume fractions of inclusions. Moreover, the instantaneous wave speed is also globally affected, being much reduced at high frequencies by the dendritic fine structure.

It is important to note that the better performance of materials with dendritic vs circular inclusions does not hold at low frequencies, where the larger volume fraction which characterizes the equivalent circular inclusions gives the major contribution to attenuation, besides resonance effects at specific frequencies. If on one side this could suggest to use dendritic inclusions with larger volume fraction for an optimized nanocomposite, it is worth reminding that the increase in the average inclusion size will translate into an increase of the affected wavelengths, and thus a decrease of the corresponding frequencies. A compromise between the frequency range that ones aims at affecting, and the extent of the attenuation needs thus to be found.

Concerning the effect of the stiffness ratio between inclusions and matrix,  for soft inclusions, the propagative contribution is reduced also at low frequencies, while for hard inclusions, it is the diffusive which is reduced at all frequencies. As such, being usually the propagative contribution dominant for thermal transport, we expect soft dendritic inclusions to represent the optimized nanostructure for affecting thermal transport. To go further on in this study and calculate the thermal conductivity, one should calculate the corresponding vibrational densities of states.
It is probably affected as well by the dispersion at the atomic scale in the Thz range. It was shown recently that an effective elasto-viscous law must be considered in this case, in place of the elastic materials behavior used in this work\cite{Luo2020,Luo2021}.

Finally, it is important to note that the wave dynamics in nanocomposites with  a complex geometrical shape such as the dendritic one is much more complex than a simple
transition from ballistic to diffusive energy transfer. Indeed, when we have shown that above
a critical frequency $\omega_c$, the nanocomposite with circular
inclusions gives rise to a clear diffusive attenuation
combined with a reduction of oscillations,
while in the same frequency range the attenuation appears to be only initially diffusive and then much stronger  for the dendritic inclusions, while presenting similar oscillations. This is different from the anomalous diffusion of acoustic waves
reported in 2D periodic media, characterized by the occurrence
of heavy-tailed distribution (as opposed to $1/x$ decay), interpreted as a consequence of the hybridization of the ballistic and diffusive transport \cite{Buonocore2019,Desmarchelier2021}. In fact, it can be fitted by a compressed exponential function with $\beta > 1$.
Therefore, this is truly another type of attenuation process due to the combined effect of periodicity and complex interface shape, giving a stronger attenuation than normal diffusion. The stretched (compressed) exponential character of this attenuation could be related for example  to the superposition of Beer-Lambert laws with competing mean-free paths due to different attenuation scales \cite{Wikipedia2021}.

All these results indicate that the use of a complex sub-structure of the interfaces in a phononic material can allow to realize novel optimized materials for acoustic attenuation, leading to applications as high-frequency acoustic filters or thermal insulation, depending on the lengthscale of the micro(nano)-structure. Indeed, all our findings, reported in normalized units, can be easily scaled at larger or smaller frequencies depending on a smaller or larger nanostructure. As such, our work is more general and gives insights on the universal effect of a complex shape onto acoustic attenuation at all lengthscales.

\begin{acknowledgments}
H.L. is financed by the french ministry of research.
V.M.G. acknowledges funding from Lyon Idex Breakthrough (IPPON) and the ANR (ANR-20-CE05-0046) for research on phonons propagation in nanocomposites.  Y.R. acknowledges NPU Short-term Overseas Visit \& Study and Innovative Experimental Project for Postgraduates.
\end{acknowledgments}

\appendix

\section{Mechanical properties of the material}
\begin{table}[ht]
\centering
\begin{tabular}{cccccccc} 
\hline
Alloy & $\sigma_y$ (MPa) &$\epsilon_y$ (\%) & $\sigma_u$ (MPa) \\
\hline
 \makecell[c]{Ti\textsubscript{45}Zr\textsubscript{25}Nb\textsubscript{6}- \\Cu\textsubscript{5}Be\textsubscript{17}Sn\textsubscript{2}}& 913 & 1.44 &1521  \\\hline \hline
 $\epsilon_u$ (\%)& $E$ (GPa) & $G$ (GPa)& $\nu$ \\
\hline
 10.12 & 85.23 $\pm$ 0.22 & 31.23 $\pm$ 0.13 & 0.365 $\pm$ 0.005 \\

\hline
\end{tabular}
\caption{Mechanical and intrinsic properties of the Ti\textsubscript{45}Zr\textsubscript{25}Nb\textsubscript{6}Cu\textsubscript{5}Be\textsubscript{17}Sn\textsubscript{2} BMG composites. Yielding strength ($\sigma_y$), yielding strain ($\epsilon_y$), ultimate tensile strength ( $\sigma_u$), tensile strain till necking ($\epsilon_u$), Young's modulus ($E$), shear modulus ($G$) and Poisson's ratio ($\nu$)
 \cite{Xu2018}.}
\end{table}    
  
\begin{table}[ht]
\centering
\begin{tabular}{cccc} 
\hline
Phase component &E (GPa) & H (GPa)\\
\hline
Dendrite-phase & 86.4 $\pm$ 4.1 & 3.78 $\pm$ 0.39\\
\hline
Glass-matrix & 113.8 $\pm$ 2.5 & 6.13 $\pm$ 0.41\\
\hline
\end{tabular}
\caption{Young's modulus (E) and   Hardness (H) of the dendrite-phase and glass-matrix in the Ti\textsubscript{45}Zr\textsubscript{25}Nb\textsubscript{6}Cu\textsubscript{5}Be\textsubscript{17}Sn\textsubscript{2} BMG composites measured by the nanoindentation. \cite{Zhai2016}.}
\end{table}  

\section{Bezier interpretation}

\begin{figure*}[htbp!]
\centering
\includegraphics[width=12cm]{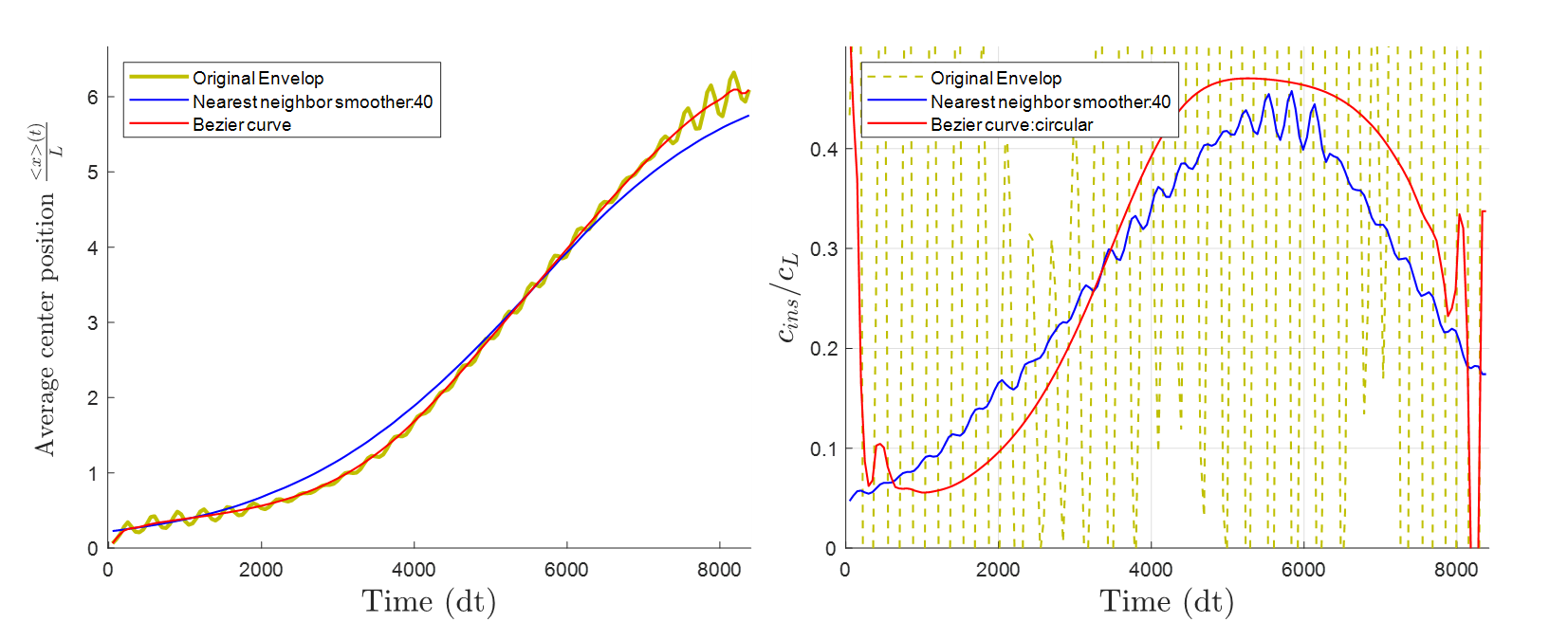}
  \caption{Average position (left)  and instantaneous speed (right) for dendritic inclusion with $E_i/E_m=0.2$ and $\frac{\omega}{\omega_0}$ = 0.6 for  longitudinal waves. Yellow lines: unsmoothed data; Blue lines: Nearest neighbors smoother; Red lines: Bezier interpolation. }
  \label{fig:dend_instantaneous_speed_1}
\end{figure*}

Before calculating the instantaneous speed, we  need to pre-treat the data of $<x>$.  Since when waves pass through a deeply heterogeneous medium, $<x>(t)$ oscillates sharply , the calculated wave speed can be unreal and meaningless, as illustrated by the yellow dashed line shown in the right panel of  Fig.\ref{fig:dend_instantaneous_speed_1}.  Two smoothing methods are considered here: the first one is \textbf{Nearest neighbor smooth} (kernel smoother) defined as:
\begin{equation}\label{eq:dend_NNS_smooth}
S_i = \frac{\sum_{j=i-n}^{i+n} P_j}{2n}
\end{equation}
where $P_j$ is the energy envelope at $j$ and  $n$ is the number of the nearest neighbors. And the second method is Bezier interpolation smoothing. \textbf{Bezier curves} can be defined for any degree $n$:
\begin{equation}\label{eq:dend_Bezier_smooth}
B(t) = \sum_{i=0}^m C_m^i (1-t)^{m-i}t^iP_i,  t\in [0,1]
\end{equation}
where $C_m^i$ equals to the binomial coefficient and   $m+1$ equals to the length of the  array $P$. It is reported that Bezier based smooth curve gives smaller fluctuations and curvatures than other regular smoothing methods in Ref. \cite{Zhou2011}, meaning that it can effectively reduce the oscillations of the first derivative of the $<x>$.

We used both methods for smoothing $<x>(t)$ for   dendritic inclusions with
$\frac{E_i}{E_m}$ = 0.2 and $\frac{\omega}{\omega_0}$ = 0.6  in which case a wave front  can  still be  identified as shown in Fig.\ref{fig:dend_snapshot} thus a well-defined sound speed should be given by a quasi-constant instantaneous speed.
 The smoothed data of  $<x>(t)$ are shown in the left panel of Fig.\ref{fig:dend_instantaneous_speed_1}  by using the two smoothing methods. The Bezier curve is clearly much closer to the real data.. In the right panel, derivative of every $<x>(t)$ shows that Bezier interpolation  gives the most stable result of wave speed while the result from the unsmoothed data is useless with such a huge oscillation and the Nearest Neighbors smoother result is still quite noisy. In addition, a plateau is observed for $t \in [8,13]$ in the case of Bezier curve, which gives  a   quasi-constant value of the wave speed (around 3500 m/s) confirming the prediction of the existence of a well-defined wave-vector. However, the very beginning and end of the Bezier curve  should be ignored, because the Bezier curve must  begin and end at given points, \textit{i.e.} endpoint interpolation property, causing a much sharper oscillation than with Nearest neighbor smoother at the two ends.
Except for those extreme points, the initial stage of acceleration before the plateau corresponds to the establishment step of the wave-packet  whose duration  depends inversely on the wave-packet frequency. Compared to the Nearest neighbor smoother, Bezier interpolation   gives  a clearer presentation and interpretation  of the instantaneous speed. Therefore, in the following work, we have chosen the Bezier curve  to smooth $<x>$ to  get the instantaneous wave speed.


%

\end{document}